# NSF Convergence Approach to Transition Basic Research into Practice

Shelby Smith, Communications and Outreach, Convergence Accelerator, NSF
Chaitanya Baru, Senior Science Advisor, Convergence Accelerator, NSF

## Abstract

The National Science Foundation Convergence Accelerator addresses national-scale societal challenges through use-inspired convergence research. Leveraging a convergence approach the Convergence Accelerator builds upon basic research and discovery to make timely investments to strengthen the Nation's innovation ecosystem associated with several key R&D priority areas and practices to include the *coronavirus disease 2019, harnessing the data revolution, the future of work, and quantum technology*. Artificial Intelligence is a key underlying theme across all of these areas.

## New Approach to Use-Inspired Research

Research is often driven by a compelling societal or scientific challenge; however, it may take the researcher community years to develop a solution. To deliver tangible solutions at a faster pace, with potentially nation-wide societal impact, the National Science Foundation (NSF) is leveraging a convergence approach to transition basic research into practice through a new capability called the *Convergence Accelerator* [NSFCA 2019a]. Using the innovation processes of *human-centered design, user discovery*, and *team science*, and integrating a *multidisciplinary research approach*, including Artificial Intelligence (AI) and machine learning, the Accelerator is making timely investments to solve high-risk societal challenges through use-inspired convergence research.

The Convergence Accelerator aligns to U.S government priorities outlined in the White House FY2020 Administration Research and Development (R&D) Budget Priorities and the President's Management Agenda [MK 2018, PMA 2019]. The strategy is to build upon basic research and discovery to strengthen the Nation's innovation ecosystem. The program is associated with several key R&D priority areas and practices including artificial intelligence, information sciences and strategic computing, security, manufacturing and medical innovation, and education and training a workforce for the 21st-century economy. The program requires public-private partnerships across sectors including, government, industry, non-profit, and academia.

## Innovation Begins Here

Launched in 2019, the NSF Convergence Accelerator is structured to support two thematic convergence research "tracks" each year. AI is a key theme in three of the first four tracks, with the fourth track being focused on Quantum Technology. Additionally, the Accelerator is also funding convergence research efforts to support solutions to bolster the national response to coronavirus disease 2019 (COVID-19).

The Convergence Accelerator employs a two-phase approach, with the research activities of the teams being proactively and intentionally managed. Phase I is a nine-month planning effort where funding of up to $750,000 is provided to further develop initial concepts; identify new team members; participate in an *innovation curriculum* that is provided by the program; and develop initial prototypes. The innovation curriculum consists of training in the areas of human-centered design, team science activities, inter-team communications, and presentation coaching—all of which are deemed essential to the success of Accelerator projects. The training in Phase I prepares the teams for success in the next phase. At the end of Phase I, teams participate in a pitch competition as well as a proposal evaluation. Selected teams from Phase I then proceed to Phase II, with potential funding up to $5 Million for 24 months, where they are expected to provide high-impact deliverables by the end of the second phase.

## Open Knowledge Networks Track

To advance the progression from data to knowledge, and to drive innovation across all areas of science and engineering by fully harnessing the power of data and AI to achieve scientific discovery and economic growth, the Convergence Accelerator released an Open Knowledge Networks (OKN) track in 2019 [NSFCA 2019b]. The Open Knowledge

Network ideas was introduced in NSF's Harnessing the Data Revolution Big Idea [HDR 2017] and discussed and refined at workshops conducted by the federal Networking and Information Technology R&D (NITRD) Big Data Interagency Working Group, resulting in a report on the topic [UCSF 2017, NITRD 2018]. The focus of the OKN projects is on exploiting publicly available datasets, especially U.S. Government and other public data. Some projects are organized by "verticals", i.e., by topic domains and sectors such as biomedicine, hazards, smart health, and court records. Other projects are organized by "horizontals", focusing on the development of software tools for data ingestion, underlying representations of facts, including geospatial and spatiotemporal information, and integrated application development environments to enable development of applications that use knowledge graphs. Examples of "vertical" and "horizontal" projects include, respectively:

**Transforming the Transparency and Accessibility of Court Records**
The U.S. court system collects detailed records, however, most of the information gathered is not publicly accessible, either because the information is behind a paywall or it is scattered across multiple systems. The Transform the Transparency and Accessibility of Court Records effort, created by Northwestern University, is developing a suite of tools to enable access to court records and to gather a better understanding of litigation data. Using machine learning, the data gathered will provide correlations and trends to inform a variety of end-users such as entrepreneurs for assessing litigation costs and risk, journalists investigating equality in outcomes, and the public becoming knowledgeable of judicial processes.

**Knowledge Network Programming System—with Application to COVID-19 Science and Economics**
Knowledge networks aid to transform data, however, building applications on top of them is difficult, time-consuming, and costly. The University of Michigan is building a Knowledge Network Programming System to allow application building ease, while also improving the quality of knowledge resources. Focused on COVID-19 science and economics, the infrastructure will assist stakeholders, such as policymakers and the medical community, in addressing the virus and the economic impact. The programming infrastructure will include an intelligent knowledge compilation layer to help programmers leverage rapidly changing knowledge networks, as well as mechanisms to debug information, share knowledge transparently, and to collect knowledge provenance metadata. The long-term goal is to establish a system that provides high accuracy knowledge with little to no human oversight.

**AI and Future of Work Track**
The NSF Big Idea on the Future of Work at the Human-Technology Frontier focuses on the workforce education and training R&D priority and leverages convergence research to understand and influence the impact of AI on the workforce and on the nature of work itself. The Convergence Accelerator track focuses on development of the human-technology partnership, the design of new technologies to augment human performance, to illuminate the emerging socio-technological landscape, understand the risks and benefits of new technologies, and to foster lifelong and universal learning. Projects in this track focus on mechanisms such as predictive AI tools, economic and labor market analyses of needed skills for future work, and education and training technologies to help connect workers to jobs of the future, support workforce training and education, and help employers develop a skilled workforce. Project examples include:

**The National Labor Exchange Research Hub**
The National Labor Exchange Research Hub (https://nlxresearchhub.org), developed by the National Association of State Workforce Agencies is the first-of-its-kind national open ecosystem of real-time and historical labor market information—connecting job seekers with employers, and cost-efficient, timely, and transparent data to researchers and practitioners developing AI tools to help predict the work of the future. The Research Hub will help plan careers, communicating to jobseekers about what skills are transferrable across industries, or how to transition from one career field to another. Employers may use the research insight and tools to improve job matching and ultimately increasing skill-based hiring. Over time the Research Hub is projected to enhance the employment and talent pipeline.

**Inclusion AI for Neurodiverse Employment**
Yearly an estimated 70,000 autistic Americans enter adulthood and an estimated 85% will be underemployed relative to their skill levels, costing the U.S. an estimated $175 billion annually. Social communications have been a factor in preventing these individuals from finding and keeping employment. *Neurodiversity* is an emerging concept which considers certain neurological differences, such as Autism, Attention Deficit Hyperactivity Disorder, Dyslexia, and others, as a natural part of the human neurocognitive variation, associated not only with impairments but also with unique strengths. The Inclusion AI for Neurodiverse Employment project at Vanderbilt University is developing AI-driven tools to mitigate these challenges and empower neurodiverse individuals to gain

meaningful employment. A suite of AI-driven technologies will be integrated within virtual environments, robotic systems, human to human interaction systems, and novel assessment tools will be designed to facilitate the creation of a talent pipeline that employs neurodiverse individuals.

### Skill-XR: Skills Training and Analytics for the Manufacturing Workforce

The manufacturing industry has experienced rapid technology changes, increasing the need to reskill the manufacturing workforce. Skill-XR, developed by Purdue University, uses AI and Extended Reality (XR) to transition the future workforce across the range from apprenticeships to real-world profitable skills. The low-cost hands-on training capabilities that will be provided will disrupt the current workforce development methods by eliminating the current need for expensive software development. Although the Skill-XR project is focusing initially on the manufacturing sector, this technology will be beneficial across multiple other industry sectors like robotics, plant operations, and construction.

### Enabling AI-Innovation via Data and Model Sharing Track

Moving forward to 2020, in March 2020 the NSF Convergence Accelerator released a new solicitation which included a track on *Enabling AI-Innovation via Data and Model Sharing*, recognizing that a critical bottleneck in making rapid progress in AI is the lack of tools, platforms, and curation protocols to enable robust and transparent sharing of data and data-driven models [NSFCA 2020a]. A total of 18 teams have been funded in Phase I, which was launched in October 2020.

### Coronavirus Disease 2019 Convergence Research

In light of the emergence and spread of the COVID-19, NSF issued a letter on April 3, 2020 to the research community for proposal ideas using the NSF Rapid Response Research funding mechanism [NSF 2020]. Related to this request, the Convergence Accelerator is funding 13 COVID-19 projects to develop tools, technologies, and techniques to support U.S. industries and sectors, the workforce, and the economy. All are applying the Convergence Accelerator's principles to provide a rapid response.

Examples of use-inspired convergence research include developing a pandemic response benefits distribution system needed to meet the high demand of unemployment insurance claims; creating an integrated knowledge graphic to help government reopen the country; addressing the shortage of medical supplies and equipment, as well as identifying alternative hospital sites; and addressing the pandemic impact on the food supply chain and transportation sector.

Lastly, a *COVID Information Commons* website (https://covidinfocommons.net) has been created to facilitate knowledge sharing and collaboration across NSF funded research efforts. Launched in July 2020, the COVID Information Commons was created to facilitate collaboration among the broad range of NSF-funded efforts addressing COVID-19-related issues [Wing 2020]. The website provides mechanisms to search for NSF COVID-19 related research projects. One of the search interfaces is based on the Lingo4G large-scale document clustering software which clusters projects based on their topics [Lingo 2020]. The CIC project has also launched a webinar series where PI's of funded efforts provide brief overviews of their project via "lightning talks".

### New Innovation Model to Make Lasting Impacts

The NSF Convergence Accelerator is young—it launched as a *pilot* program in 2019 and is now only in its second year of implementation. The program is unique to NSF. It has envisioned an innovation process that provides a positive, nurturing environment for convergence research with a curriculum whose objective is to assist multi-organization teams to unify towards their own project objectives, thereby resulting in high-impact solutions for society at scale. NSF recently issued a Dear Colleague Letter requesting information on future topics for the Convergence Accelerator [NSFCA 2020b]. To engage or learn more about the Convergence Accelerator, visit https://www.nsf.gov/od/oia/convergence-accelerator/index.jsp.

## References

[HDR 2017] NSF Harnessing the Data Revolution, https://www.nsf.gov/cise/harnessingdata/.

[Lingo 2020] Lingo4G Clustering Engine, https://carrotsearch.com/lingo4g/.

[MK 2018] Mulvaney and Kratsios 2018. Memorandum for the Heads of Executive Departments and Agencies. *FY2020 Administration Research and Development Budget Priorities,* Washington, D.C.

[NITRD 2018] Open Knowledge Network: Summary of the Federal NITRD Big Data IWG Workshop, November 20, 2018, https://www.nitrd.gov/pubs/Open-Knowledge-Network-Workshop-Report-2018.pdf.

[NSF 2020] NSF 20-052, Dear Colleague Letter on the Coronavirus Disease 2019 (COVID-19), https://www.nsf.gov/pubs/2020/nsf20052/nsf20052.jsp.

[NSFCA 2019a] NSF Convergence Accelerator, https://www.nsf.gov/od/oia/convergence-accelerator/index.jsp.

[NSFCA 2019b] NSF 19-050 Dear Colleague Letter: NSF Convergence Accelerator Pilot (NSF C-Accel), https://www.nsf.gov/pubs/2019/nsf19050/nsf19050.jsp.

[NSFCA 2020a] NSF 20-565 NSF Convergence Accelerator Phase I and II, https://www.nsf.gov/pubs/2020/nsf20565/nsf20565.htm.

[NSFCA 2020b] RFI NSF 21-012 Request for Information on Future Topics for the NSF Convergence Accelerator, https://www.nsf.gov/pubs/2021/nsf21012/nsf21012.jsp.

[PMA 2019] *President's Management Agenda,* Washington, D.C.: Performance.gov.

[UCSF 2017] Third Workshop on an Open Knowledge Network: Enabling the Community to Build the Network, Oct 4-5, National Library of Medicine, Bethesda, MD, Oct 2017, https://bakarinstitute.ucsf.edu/open-knowledge-network/.

[Wing 2020] COVID Information Commons (CIC), PI: Jeannette Wing, Columbia University, https://www.nsf.gov/awardsearch/showAward?AWD_ID=2028999&HistoricalAwards=false